\documentclass[prl,twocolumn,showpacs,preprintnumbers,amsmath,amssymb,superscriptaddress]{revtex4}
\usepackage{bm}
\usepackage{epsfig,amsopn}
\usepackage{graphicx}
\usepackage{amsmath,amssymb}
\usepackage{natbib}
\renewcommand{\section}[1]{{\par\it #1.---}}
\def\be{\begin{equation}}
\def\ee{\end{equation}}
\def\bea{\begin{eqnarray}}
\def\eea{\end{eqnarray}}
\def\la{\langle}
\def\ra{\rangle}

\def\nn{\nonumber}

\def\f{\frac}

\def\etal{{\emph{et al~}}}

\def\ie{{\emph{i.e~}}}

\def\bn{{\bf{n}}}
\def\be{{\bf{\hat{e}}}}

\begin{document}

\title{ Heat conduction in  a three dimensional anharmonic crystal }
\author{Keiji Saito}
\email{saitoh@spin.phys.s.u-tokyo.ac.jp}
\affiliation{Graduate School of Science, 
University of Tokyo, 113-0033, Japan} 
\affiliation{CREST, Japan Science and Technology (JST), Saitama, 332-0012, Japan}
\author{Abhishek Dhar}
\email{dabhi@rri.res.in}
\affiliation{Raman Research Institute, Bangalore 560080, India}
\date{\today} 
\begin{abstract}
We perform nonequilibrium simulations of heat conduction in a three
dimensional anharmonic lattice. By studying slabs of length $N$
and width $W$, we examine the cross-over from one-dimensional to  three 
dimensional behavior  of the thermal conductivity
$\kappa$. We find that for large $N$, the cross-over takes place at a
small value of the aspect ratio $W/N$.   
>From our numerical  data  we conclude that the three dimensional system has a
finite non-diverging $\kappa$ and thus provide the first verification of
Fourier's law in a system without pinning.   
\end{abstract}

\maketitle
Macroscopic behavior of heat transport in the 
linear response regime is governed by Fourier's law
\bea
\bar{J} &=& -\kappa \bar{\nabla} T(\bar{x}) , \label{fourier}
\eea
where ${\bar{J}}$, $\bar{\nabla}T$ are respectively the heat current
density and temperature 
gradient at the position $\bar{x}$, and $\kappa$ is the thermal conductivity.
This implies diffusive behavior of heat. 
What are the necessary and sufficient conditions for the validity
of Fourier's law ? This question is a longstanding unsolved problem \cite{BLR00}. 
For solids one starts with the
description in terms of a harmonic crystal where 
heat conduction takes place through lattice vibrations or phonons. Scattering 
of the phonons can occur due to phonon-phonon interactions 
({\emph{i.e}}~ anharmonicity in the interactions) 
or by impurities (e.g~ isotopic disorder, defects) \cite{ziman72}.
For one dimensional systems, from a large number of numerical and
analytical studies it is  
now established that these scattering mechanisms are
insufficient in ensuring normal diffusive transport. 
Instead one finds anomalous transport \cite{dhar08,LLP03}, one of the
main signatures of this being that the thermal conductivity 
$\kappa$ in such systems is no longer an intrinsic material property but 
depends on the linear size $N$ of the system. A power law dependence $\kappa
\sim N^\alpha$ is typically observed. For two dimensional anharmonic
crystals a $\kappa \sim \ln(N)$ divergence of the conductivity is predicted
from various analytical theories \cite{llp98,NR02} and also from an exactly
solved stochastic model \cite{basile06}, but the numerical
evidence for this so far is inconclusive \cite{LL00,grassyang02}.   
A recent experiment has reported the
breakdown of Fourier's 
law in nanotubes \cite{chang08} while another experiment on graphene flakes
\cite{nika09} also indicates a divergence of $\kappa$.

For systems with pinning (\ie~an external substrate potential) and
anharmonicity, Fourier's law has been verified in simulations on one
and two dimensional systems \cite{dhar08}. 
There is a strong belief that Fourier's law should be valid in three
dimensional ($3D$) systems, even without pinning. A recent work
\cite{chaudhuri09} examined heat transport in a $3D$ disordered
harmonic crystal. Analytical arguments showed that heat conduction in the system
was sensitive to boundary conditions. For generic boundary conditions
a finite conductivity was predicted but this could be numerically
verified only for the pinned case.
In this letter we investigate the effect of anharmonicity on heat conduction
in ordered  crystals. Through extensive simulations of a $3D$
anharmonic crystal 
we give strong numerical evidence for normal transport and the
validity of Fourier's law in this system.   

\section{Model}
We consider a $3D$ cubic crystal  with a scalar displacement field
$x_\bn$ defined on 
each lattice site $\bn=(n_1,n_2,n_3)$ where $n_1=1,2,...,N$ and
$n_2=n_3=1,2,...,W$. The Hamiltonian is taken to be of the
Fermi-Pasta-Ulam (FPU) form:
\bea
H&=&\sum_{\bn} \f{{\dot{x}}_\bn^2}{2}+ \sum_{\bn,\be} [~\f{1}{2} (x_\bn-x_{\bn+\be})^2
+ \f{\nu}{4} (x_\bn-x_{\bn+\be})^4 ]\, , ~ \nonumber \\
\label{ham}
\eea
where $\be$ denotes unit vectors in the three directions.
We have set the values of all masses and harmonic spring constants to
one and the 
anharmonicity parameter is $\nu$. Two of the faces of the crystal, namely those
at $n_1=1$ and 
$n_1=N$, are coupled to white noise Langevin type heat baths so that the equations of
motion of the particles are given by:
\bea
\ddot{x}_\bn&=&-\sum_\be [~(x_\bn-x_{\bn+\be}) + \nu (x_\bn-x_{\bn+\be})^3~ ] \nn
\\ 
&+&  \delta_{n_1,1} (-\gamma \dot{x}_\bn+\eta^L_{\bn}) +  \delta_{n_1,N}
(-\gamma \dot{x}_\bn+\eta^R_{\bn})~. \label{eqm} 
\eea 
The noise terms at different sites are uncorrelated while at a given site the
noise strength is specified by $\la \eta^{L,R}_{\bn}(t)
\eta^{L,R}_{\bn}(t') \ra = 2 \gamma T_{L,R} \delta (t-t')$ , where $T_L$
and $T_R$ are the temperatures of the left and right baths and we have chosen
units where  the Boltzmann constant $k_B=1$. Fixed boundary conditions
were used for the particles connected to the baths and periodic
boundary conditions were imposed in all the other directions.
We simulate these equations using a velocity-Verlet algorithm \cite{AT87} and
calculate the  heat current and the temperature profile in the nonequilibrium
steady state of the crystal. The heat current $j_\bn$ from the lattice site
$\bn$ to $\bn+\be_1$ where $\be_1=(1,0,0)$,  is given by $j_\bn=\la
f_{\bn,\bn+\be_1} \dot{x}_{\bn+\be_1} \ra$, with $f_{\bn,\bn+\be_1}$ being the
force on the particle at site $\bn+\be_1$ due to the particle at site $\bn$. 
In our simulations we calculate the average current per bond given by 
\bea
J=\f{1}{{W}^2(N-1)} \sum_{n_1=1}^{N-1} \sum_{n_2,n_3=1}^{W} j_\bn~. \nn 
\eea
We also calculate the average  temperature across  layers in the slab and this
is given by $T_{n_1}=(1/{W}^2)\sum_{n_2,n_3} \dot{x}_\bn^2$.

\section{Simulation details} 
\begin{figure}
\vspace{1cm}
\includegraphics[width=3.2in]{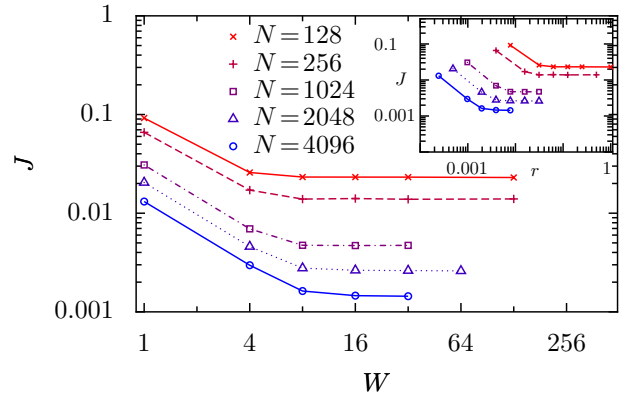}
\caption{Plot of the heat current $J$ versus width $W$ for different fixed
  values of the length $N$. The inset plots $J$ versus the aspect ratio $r=W/N$.}
\label{crossover}
\end{figure}

In all our simulations we set $\nu=2$ and $T_L=2,
T_R=1$. We first address the question of the dependence of $J$ on the
width $W$ of the system and the nature of the cross-over from $1D$ behaviour,
for small values of the ratio $r=W/N$, to true $3D$ behavior for $W/N \sim
1$. The numerical results are given in Fig.~(\ref{crossover}). We see that for any fixed
length $N$, the value of $J$  decreases as we increase $W$ but saturates
quickly to the $3D$ value. The cross-over width $W_c$ is seen to
increase slowly with $N$. The inset shows that as we increase $N$, the
cross-over from  $1D$  to $3D$ behavior takes  place at decreasing
values of  $r$ and presumably in the thermodynamic limit $N \to
\infty$, the cross-over occurs at $r \to 0$. 
Thus our study suggests that $W_c \sim N^a$ with $0<a<1$. 
A similar result was
obtained by Grassberger and Yang \cite{grassyang02} 
for a $2D$ FPU system. 

\begin{figure}
\vspace{1cm}
\includegraphics[width=3.2in]{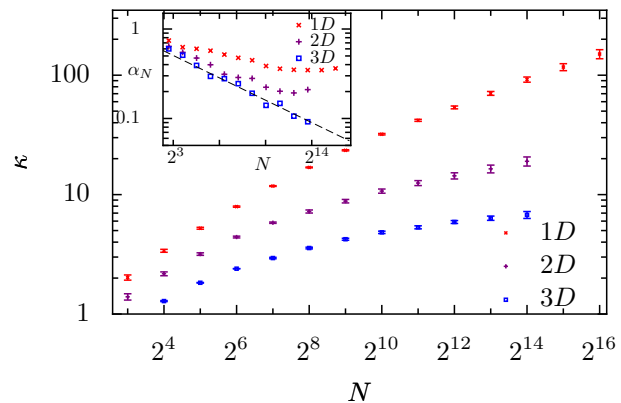}
\caption{Plot of $\kappa$-versus-$N$ in different dimensions. The inset shows
  the running slope $\alpha_N=d\ln {\kappa}/ d \ln{N}$ as a function of $N$.
The dashed line is a guide to the eyes.
}  
\label{jvsn}
\end{figure}

Next we look at the   dependence of $J$ on $N$ for the $3D$ case. The fast
cross-over from $1D$ to $3D$ behaviour implies that we can extrapolate the 
results for small $r$ to estimate the true value of the $3D$ current (at
$r=1$). Thus we can get 
results for quite large values of $N$ from  simulations on systems with 
small widths. For sizes up to $N=128$ we obtained data for $W=N$.
For the largest system size, namely $N=16384$ we have data for
$W=16$.  We show our results for the $N$ dependence of $\kappa$ in
Fig.~(\ref{jvsn}).  
There are three sources of error in the values of current: (i) numerical
errors, arising from the finite time discretization value 
($dt=0.001$), and from rounding off errors; (ii) statistical errors
arising from averaging over a finite number of time steps; and (iii)
errors arising from the extrapolation of the small aspect ratio ($r$)
results to the $3D$ case. The error from (iii) was taken to be the
difference in current values for the two largest widths studied. For
smaller system sizes we verified  that the numerical error was much
smaller than the statistical and extrapolation errors and we assume
that this is true also at larger system sizes. The error-bar for each
data point plotted in Fig.~(\ref{jvsn}) is  the larger of errors from
(ii) and (iii). 

The slope of the 
$\kappa$ versus $N$ curve is decreasing slowly with $N$ and a straight line fit
to the last three points  gives an exponent $\alpha=0.09 \pm 0.01$. 
For comparison we also show in Fig.~(\ref{jvsn}) the $1D$ and $2D$
data for the FPU system. The $2D$ results are from data for $N \times N$
samples for 
systems up to $N=2048$ while for larger sizes the results shown are
extrapolated values from small width data. In $1D$  we get  $\alpha
\approx 0.33$ \cite{mai07} while in $2D$ we get $\alpha \approx 0.22$.    
In the inset of Fig.~(\ref{jvsn}) we have plotted the running slope
defined as $\alpha_N=d\ln {\kappa}/ d \ln{N}$ against system
size. From this we see that while the slopes in $1D$ and $2D$ tend to
saturate, the $3D$ slope seems to be decreasing. 
%%KS 
The $3D$ slope 
can be fitted by the dashed line with a  power law form.
This suggests that
the asymptotic system size behaviour will  give $\alpha=0$
implying diffusive transport and validity of Fourier's law. 

\begin{figure}
\vspace{1cm}
\includegraphics[width=3in]{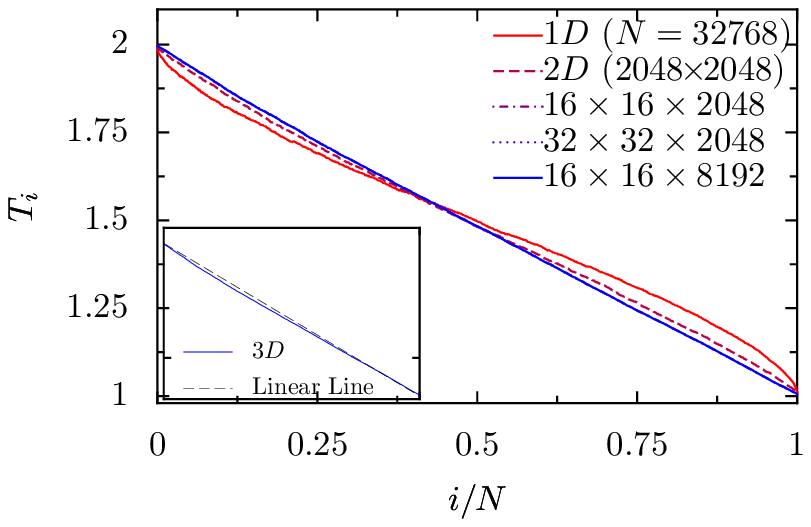}
\caption{Plot of temperature profiles for a $1D$ system, a $N \times N$ $2D$
 system and a $W\times W\times N$ $3D$ system with different aspect
  ratios  $r=W/N$. 
%%KS
Temperature profiles for the three aspect ratios
overlap with each other. The inset shows that the $3D$ temperature profile
 is concave upward everywhere.}  
\label{temp_pro}
\end{figure}

One of the remarkable features of $1D$ systems with anomalous
heat transport is the form of the steady state temperature profile
obtained in these  
systems. Typically one finds that the temperature profile is concave
upwards in part of the system and concave downwards
elsewhere   and this is true even for small temperature differences
\cite{LL00,mai07,LMP09}. This means that the temperature gradient is
non-monotonic as a function of distance across the sample.  
 In Fig.(\ref{temp_pro}) we plot the temperature
profiles for the $1D$, $2D$ and $3D$ samples. We see that the 
variation of the temperature gradients are
non-monotonic in both $1D$ and $2D$ while in $3D$ they are
monotonic. 
The inset in Fig.~(\ref{temp_pro})shows that the $3D$ temperature
profile is concave upward everywhere. We have also confirmed that the
profile becomes more linear on decreasing the temperature difference
between $T_L$ and $T_R$.  
This again supports our finding based on the size-dependence of the current, 
that heat transport in $3D$ is diffusive while in lower dimensions it
is anomalous.  

\begin{figure}
\vspace{1cm}
\includegraphics[width=3in]{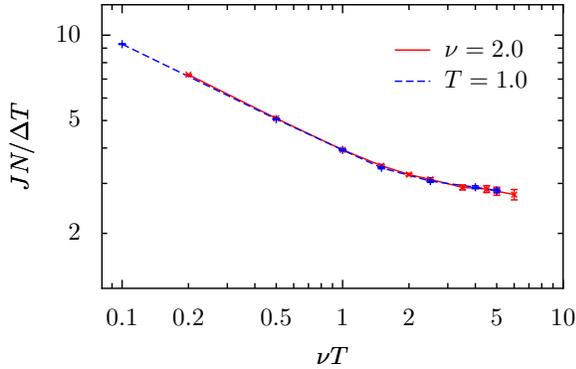}
\caption{Demonstration of scaling property (\ref{conductivity_scaling}) 
for $32 \times 32 \times 128$ system with $\Delta T=0.1$. 
Thermal conductivities decrease as increasing temperature $T$ 
or nonlinearity $\nu$. 
This temperature dependence of $\kappa$ explains
the slightly concave curve in temperature profiles in $3D$ (see Fig.(\ref{temp_pro})).}
\label{temp_depe}
\end{figure}

Finally we look at the temperature dependence of thermal conductivity. 
Temperature and nonlinearity are highly correlated \cite{AL08}, and  
temperature dependence can be understood from the nonlinearity dependence of 
thermal conductivity. 
We note that Eq.~(\ref{eqm}) leads to the scaling relation 
$s J(T,\Delta T , s \nu)=J(s T , s \Delta T , \nu) $, where $T=(T_L + T_R)/2$,
$\Delta T = T_L - T_R$, and $s$ is an arbitrary scale factor.
Taking the limit $\Delta T\to 0$, this gives 
the scaling relation for thermal conductivity as 
$\kappa (T, s \nu)=\kappa (sT , \nu )$. Putting $\nu = 1$ and $s = \nu$, 
we then get 
\bea
\kappa(T, \nu) = \kappa (\nu T , 1) \label{conductivity_scaling} .
\eea
Thus the thermal conductivity is a function of $\nu T$. One may expect that 
large $\nu$ suppresses heat currents due to enhancement of 
phonon-phonon interactions. Hence from the scaling
(\ref{conductivity_scaling}) we expect that  
$\kappa$ must also decrease with increasing $T$.  
To check this, we show the dependence of the heat 
current on $\nu T$ for a $32 \times 32 \times 128$ 
system with a small temperature difference $\Delta T=0.1$.
In Fig.(\ref{temp_depe}), we compared two cases: 
one with $\nu=2.0$ fixed and $T$  varied, 
and another with $T=1.0$ fixed and $\nu$  varied.
We find that current  decreases as a function of  $\nu T$, consistent
with the scaling relation  Eq.~(\ref{conductivity_scaling}).
We note that Fourier's law (\ref{fourier}) leads to 
$d^2 T/ d \bar{x}^2 = -J^2 \kappa^{-3} (d\kappa/dT)$ and so  
the decrease of $\kappa$ in the region $T \in[1.0,2.0]$ 
with $\nu=2.0$ is consistent with the concave curve in the $3D$ temperature
profiles. Interestingly at large anharmonicity the current does not
seem to go to zero but instead appears to saturate to
a constant value. At low temperatures the effect of anharmonicity becomes
weaker and we expect the conductivity  to increase, eventually diverging in
the limit $T \to 0$. It is difficult to numerically access the low temperature
regime since the mean free path becomes large and one would need much larger
system sizes to see  diffusive behaviour.

\section{Summary and Discussion}
In summary, we have given the first numerical evidence for the
validity of Fourier's law of heat conduction in an anharmonic crystal
in three dimensions. This confirms the belief that in three dimensions anharmonicity is a
{\emph{sufficient}} condition for normal transport. This is not a necessary condition
since, for example,  a $3D$ pinned disordered purely harmonic crystal also shows normal transport \cite{chaudhuri09}.
Our conclusion was based on three evidences. The first is
the system-size dependence of the thermal conductivity, the second is 
temperature profile, and the third is the consistency between temperature 
profile and temperature dependence of conductivity.
It has been known that the one-dimensional FPU system shows
slow convergence of the thermal conductivity to it's asymptotic
behavior \cite{mai07}. 
Here we show that this is also the case  in $3D$. Unlike $1D$ and $2D$,
the running slope of the 
size dependence of $\kappa$ in $3D$ showed decreasing behavior even 
at the largest system size and this gives us a clear signature for
finite $\kappa$.   
The temperature profiles in $3D$ are completely different type from
the $1D$ and $2D$ case where nonmonotonic behavior of the gradient is
robust even for small temperature  
differences. 
We note that a recent simulation of heat conduction in the $3D$ FPU crystals
reported diverging thermal conductivity 
(the reported exponent is about $0.221$) \cite{shiba08}. The reasons for this is
probably because of the small values of anharmonicity used in those
simulations and also the much smaller 
system sizes that were studied (maximum size in that study was $N=256$).
In $2D$ we find a divergence of the
conductivity with an exponent $\alpha \approx 0.22$ which is similar to
the value obtained in \cite{grassyang02}.

For a sample of fixed length $N$ we find that the
current density decreases on increasing its width $W$ and 
the cross-over from $1D$ to $3D$ behaviour takes place at a value $W_c
\sim N^a$ with $0<a<1$. 
This has  implications for experiments measuring thermal conductivity
of  nanowires \cite{tighe97,schwab00,liwu03}. 
If the cross-over width were independent of $N$ and 
the width of the nanowire larger than it, then the thermal conductivity of long
nanowires could well be finite and not diverge as expected for true
$1D$ systems.    
On the other hand, since the cross-over width gradually increases 
with increasing $N$, a gradual transition from $3D$-like to $1D$
behavior  will take place when the cross-over width is comparable to
the width of nanowires.  This scenario is an interesting system size
effect that may be observed in experiments on nanowires.  

We thank H. Shiba, N. Ito, and N. Shimada for useful discussions and showing us
unpublished data on  auto-correlation functions of heat currents.
KS was supported by MEXT, Grant Number (21740288).

\begin{thebibliography}{} 

\bibitem{BLR00}
F. Bonetto, J.L. Lebowitz, and L. Rey-Bellet, in {\em Mathematical Physics
2000}, edited by A. Fokas et. al. (Imperial College Press, London, 2000), 
p. 128.

\bibitem{ziman72}  J. M. Ziman, {\it Principles of the Theory of
  Solids},(Cambridge University Press, Cambridge, 1972).

 
\bibitem{dhar08} A. Dhar,   Adv. Phys. {\bf 57}, 457 (2008). 

\bibitem{LLP03} S. Lepri, R. Livi, and A. Politi, Phys. Rep. {\bf 377}, 1
  (2003). 






\bibitem{llp98} S. Lepri, R. Livi and A. Politi,
  Euro. phys. Lett. {\bf 43}, 271 (1998). %% MCT refn. 

\bibitem{NR02} O. Narayan and S. Ramaswamy, Phys. Rev. Lett. {\bf
  89}, 200601 (2002).

\bibitem{basile06} G. Basile, C. Bernardin, S. Olla, 
  Phys. Rev. Lett. {\bf 96}, 204303 (2006).

\bibitem{LL00} A. Lippi and R. Livi, J. Stat. Phys. 100, 1147
  (2000). %% 2D FPU

\bibitem{grassyang02} P. Grassberger and L. Yang, cond-mat/0204247.


\bibitem{chang08} C. W. Chang \etal~, 
Phys. Rev. Lett. {\bf 101}, 075903 (2008). 

\bibitem{nika09} D.L. Nika \etal~, Appl. Phys. Lett. {\bf 94}, 203103 (2009).

\bibitem{chaudhuri09} A. Chaudhuri \etal~, arXiv:0902.3350 (2009).


\bibitem{AT87} M. P. Allen and D. L. Tildesley, {\it Computer
  Simulations of Liquids} (Clarendon, Oxford, 1987).

\bibitem{mai07} T. Mai, A. Dhar and O. Narayan, Phys. Rev. Lett. {\bf 98}, 
184301 (2007).

\bibitem{LMP09} S Lepri, C Mejia-Monasterio and A Politi, J. Phys. A
 {\bf 42}, 025001 (2009).


\bibitem{AL08} A. Dhar and J. L. Lebowitz, Phys. Rev. Lett. {\bf 100}, 134301 
(2008).

\bibitem{shiba08} H. Shiba and N. Ito, J. Phys. Soc. Jpn. {\bf 77}, 054006
  (2008). 

\bibitem{tighe97} T. S. Tighe, J. M. Worlock, M. L. Roukes,
Appl. Phys. Lett. {\bf 70}, 2687 (1997). 
 

\bibitem{schwab00} K. Schwab, E. A. Henriksen, J. M. Worlock and
  M. L. Roukes,  Nature {\bf 404}, 974 (2000).   

\bibitem{liwu03} D. Li \etal~, Appl. Phys. Lett. {\bf 83}, 2934 (2003).

\end{thebibliography}
\end{document}